\documentclass{aastex701}
\usepackage{xcolor}
\newcommand{\os}{\textsc{OrbitSift}}
\newcommand{\cmd}[1]{\texttt{#1}}
\newcommand{\repourl}{the author on request (public release planned)}
\newcommand{\nTiles}{14,932}
\newcommand{\nReq}{17,824}
\newcommand{\nTilesRows}{2,063}
\newcommand{\nDia}{7,186,991}
\newcommand{\nSel}{3,055,500}
\newcommand{\nNights}{91}
\newcommand{\firstNight}{January~1}
\newcommand{\lastNight}{November~15}
\newcommand{\nTilesBeta}{160}
\newcommand{\nDeep}{65}
\newcommand{\nObj}{241}
\newcommand{\nObjOffset}{228}
\newcommand{\nObjLinkOnly}{13}
\newcommand{\nObjXfive}{223}
\newcommand{\nRowsAll}{1,914}
\newcommand{\nNew}{1,616}
\newcommand{\nNewNights}{886}
\newcommand{\latestXfive}{July~10}
\newcommand{\offMed}{10.0}
\newcommand{\offGtOne}{209}
\newcommand{\offGtFive}{164}
\newcommand{\nTab}{237}
\newcommand{\gainMed}{8.8}
\newcommand{\gainGtTen}{110}
\newcommand{\gainGtHundred}{29}
\newcommand{\nGainObj}{229}
\newcommand{\rmsMax}{0.33''}
\newcommand{\oursMax}{0.75}
\newcommand{\wBefore}{1.97}
\newcommand{\wAfter}{0.18}
\newcommand{\arcBeforeAl}{59}
\newcommand{\arcAfterAl}{378}
\newcommand{\nBigA}{16}
\newcommand{\magMed}{22.8}
\newcommand{\magLo}{21.6}
\newcommand{\magHi}{24.0}
\newcommand{\nLinksWide}{118}
\newcommand{\nLinksWideId}{100}
\newcommand{\nLinksWideUn}{18}
\newcommand{\nLinksWideUnFour}{15}
\newcommand{\nLinksWideUnNoorb}{4}
\newcommand{\nCtlWide}{13.0}
\newcommand{\nCtlWideTot}{39}
\newcommand{\ctlTopTile}{18}
\newcommand{\nCtlWideNoorb}{18}
\newcommand{\rmsIdHi}{0.20}
\newcommand{\rmsCtlLo}{0.17}
\newcommand{\rmsCtlHi}{0.51}
\newcommand{\ctlDaysLo}{15}
\newcommand{\ctlDaysHi}{54}
\newcommand{\idDaysLo}{5}
\newcommand{\idDaysHi}{27}
\newcommand{\pGe}{0.11}
\newcommand{\recallN}{76}
\newcommand{\recallOf}{77}
\newcommand{\recallLo}{93}
\newcommand{\recallHi}{99.8}
\newcommand{\nDeepReal}{992}
\newcommand{\nDeepRealFour}{981}
\newcommand{\nDeepCtl}{936}
\newcommand{\nDeepId}{8}
\newcommand{\nDeepUnFive}{4}
\newcommand{\nDeepCtlFive}{3}
\newcommand{\posN}{400}
\newcommand{\posIn}{none}
\newcommand{\nPop}{2,485}
\newcommand{\nPopA}{2,424}
\newcommand{\nPopC}{61}
\newcommand{\nXfiveAll}{2,387}
\newcommand{\nXfiveC}{32}
\newcommand{\pXfive}{98.5}
\newcommand{\checknorbits}{9,375}
\newcommand{\nBodies}{240}
\newcommand{\nNoneAtMPCbody}{219}
\newcommand{\nGrpArc}{358}
\newcommand{\nGrpAfter}{287}
\newcommand{\nGrpBefore}{971}
\newcommand{\nBefJunJul}{838}
\newcommand{\nBefEarly}{133}
\newcommand{\nBefRubinTwentyfive}{809}
\newcommand{\nBefXtwentyfive}{823}
\newcommand{\nBefXtwentysix}{46}
\newcommand{\nBefOther}{102}
\newcommand{\nCatAll}{11}
\newcommand{\nNewEarlyFeb}{233}
\newcommand{\bThreeRms}{0.09}
\newcommand{\bThreeMax}{0.27}
\newcommand{\bThreeA}{46.89}
\newcommand{\bThreeAsig}{0.02}
\newcommand{\bThreeE}{0.177}
\newcommand{\bThreeI}{7.57}
\newcommand{\bThreeUsed}{21}
\newcommand{\bThreeCount}{21}
\newcommand{\bThreeOld}{4}
\newcommand{\bThreeNew}{17}
\newcommand{\unidFourLow}{5}
\newcommand{\unidFourLowRmsLo}{0.01}
\newcommand{\unidFourLowRmsHi}{0.08}
\newcommand{\unidFourLowDaysLo}{6}
\newcommand{\unidFourLowDaysHi}{18}
\newcommand{\unidFourRest}{10}
\newcommand{\unidFourRestRmsLo}{0.30}
\newcommand{\unidFourRestRmsHi}{0.49}
\newcommand{\unidFourRestDaysLo}{31}
\newcommand{\unidFourRestDaysHi}{54}
\newcommand{\recallDeepN}{8}
\newcommand{\recallDeepOf}{11}
\newcommand{\gainLtOneFive}{23}
\newcommand{\gainLtOne}{3}
\newcommand{\posObj}{517}
\newcommand{\posRows}{1,501}
\newcommand{\posInSeason}{57}
\newcommand{\posDropTime}{7}
\newcommand{\posSnrLtSeven}{349}
\newcommand{\posSnrLtEight}{581}
\newcommand{\adesEightObj}{300}
\newcommand{\adesEightRows}{769}
\newcommand{\adesEightSec}{259}
\newcommand{\nProps}{132}
\newcommand{\nNewMps}{905}
\newcommand{\nNewNoref}{711}
\newcommand{\nMpsObj}{153}
\newcommand{\nMpsObjX}{140}
\newcommand{\nMpsObjOther}{13}
\newcommand{\nNorefObj}{87}
\newcommand{\mpsLbDate}{March~25}
\newcommand{\nMpsBefore}{665}
\newcommand{\nMpsAfter}{240}
\newcommand{\nNorefArc}{358}
\newcommand{\nNorefBefore}{306}
\newcommand{\nNorefAfter}{47}
\newcommand{\nMpsLate}{847}
\newcommand{\mpsLateOff}{8.5}
\newcommand{\pMpsLateGtOne}{94}
\newcommand{\nMpsLateObj}{150}
\newcommand{\pMpsShare}{52}
\newcommand{\nBodiesXfive}{222}
\newcommand{\bPredOffLo}{9.7}
\newcommand{\bPredOffHi}{9.8}
\newcommand{\bPredSd}{14}
\newcommand{\bPredRms}{0.10}
\newcommand{\bPredA}{46.73}
\newcommand{\bPredAsig}{0.21}
\newcommand{\bNorejUsed}{21}
\newcommand{\bNorejCount}{21}
\newcommand{\bNorejRms}{0.09}
\newcommand{\bNAlert}{5}
\newcommand{\eScreenOld}{5,574}
\newcommand{\eScreenTol}{1}
\newcommand{\eScreenCand}{473}
\newcommand{\eIdent}{4}
\newcommand{\ePossible}{3}
\newcommand{\eIdentList}{2026~GT$_{27}$ = 2003~GM$_{53}$, 2026~EP$_{148}$ = 2016~FH$_{59}$, 2026~EY$_{127}$ = 2022~GN$_{14}$, 2025~NK$_{497}$ = 2002~PE$_{153}$}
\newcommand{\ePossList}{2025~MN$_{387}$ and 2000~JH$_{81}$, 2025~MW$_{350}$ and 2004~LZ$_{32}$, 2025~ML$_{352}$ and 2004~MM$_{10}$}
\newcommand{\eIdentAll}{5}
\newcommand{\densMedNight}{105}
\newcommand{\offMedDet}{2.6}
\newcommand{\offMedNew}{4.5}
\newcommand{\mpsLateCat}{173}
\newcommand{\mpsLateLeOne}{55}
\newcommand{\mpsRefLo}{September~1}
\newcommand{\mpsRefHi}{September~16}
\newcommand{\bPredCross}{0.1}
\newcommand{\bPredCrossSd}{2}
\newcommand{\bPredAlongSig}{0.7}
\newcommand{\bPredStrip}{27}
\newcommand{\posOneRow}{3}
\newcommand{\posOneRowList}{2026~EP$_{148}$, 2026~EU$_{123}$, 2026~GR$_{25}$}
\newcommand{\eExpTenth}{4.7}
\newcommand{\eExpHundredth}{0.05}
\newcommand{\eWithinHundredth}{7}
\newcommand{\eChanceTenth}{3}
\newcommand{\eCtlN}{20}
\newcommand{\eCtlIdent}{0}
\newcommand{\eCtlPossible}{0}
\newcommand{\qkArc}{27}
\newcommand{\qkOff}{1.6}
\newcommand{\nArcOnly}{8}
\newcommand{\deepCtlFivePer}{1}
\newcommand{\fetchHours}{6.0}
\newcommand{\nWideBefore}{15}
\newcommand{\wideArcLo}{7}
\newcommand{\wideArcHi}{41}
\newcommand{\nImgCent}{5}
\newcommand{\ePredN}{2}
\newcommand{\ePredCtl}{4}
\newcommand{\ePredMin}{9}
\newcommand{\ePredMax}{28}
\newcommand{\ePredPos}{0.7}
\newcommand{\uPairs}{798}
\newcommand{\uFits}{6}
\newcommand{\uNoBound}{4}
\newcommand{\uCtlN}{6}
\newcommand{\aCmpN}{487}
\newcommand{\aCmpWide}{5}
\newcommand{\aCmpArcHi}{41}
\newcommand{\aCmpGain}{96}
\newcommand{\nSearched}{2,415}
\newcommand{\nStrong}{771}
\newcommand{\nStrongClean}{649}
\newcommand{\nImgConf}{517}
\newcommand{\nImgDone}{771}
\newcommand{\arcBefore}{71~days}
\newcommand{\arcAfter}{10.1~years}
\newcommand{\nCrowded}{72}
\newcommand{\nWeak}{508}
\newcommand{\nNone}{1,064}
\newcommand{\nStrongCtl}{102}
\newcommand{\nStrongInc}{20}
\newcommand{\confRate}{71}
\newcommand{\nConfDone}{550}
\newcommand{\nFits}{517}
\newcommand{\sigGainMed}{99}
\newcommand{\sigAfterMed}{0.003}
\newcommand{\nAoffLike}{37}
\newcommand{\nUnbBefore}{13}
\newcommand{\nBoundBefore}{504}
\newcommand{\nAoffCat}{48}
\newcommand{\nAoffCatU}{47}
\newcommand{\nAoffSel}{6}
\newcommand{\nSelTargets}{13}
\newcommand{\nBadFit}{7}
\newcommand{\failSnr}{376}
\newcommand{\failFrame}{189}
\newcommand{\failCtl}{133}
\newcommand{\failStatic}{57}
\newcommand{\nCopies}{3,042}
\newcommand{\nCopiesStrong}{127}
\newcommand{\pCopy}{4.2}
\newcommand{\nCopiesThree}{123}
\newcommand{\nCopiesFour}{4}
\newcommand{\pCopyThree}{4.0}
\newcommand{\pCopyFour}{0.13}
\newcommand{\nEval}{2,343}
\newcommand{\nKzeroThree}{227}
\newcommand{\nKzeroFour}{544}
\newcommand{\nNoFour}{1,799}
\newcommand{\expChanceThree}{73}
\newcommand{\expChanceFour}{3}
\newcommand{\nCtlImg}{127}
\newcommand{\nCtlImgRows}{312}
\newcommand{\nCtlImgPass}{20}
\newcommand{\nCtlImgConf}{0}
\newcommand{\rcUl}{2.3}
\newcommand{\expFalseThreeUl}{1.7}
\newcommand{\expFalseFourUl}{0.1}
\newcommand{\nSecure}{396}
\newcommand{\nProbable}{121}
\newcommand{\nDropBtwo}{9}
\newcommand{\nDropBone}{17}
\newcommand{\nThreeCounted}{85}
\newcommand{\nThreeNotCounted}{142}
\newcommand{\expFalseThreeCons}{3.3}
\newcommand{\candsReal}{180}
\newcommand{\candsCtlLo}{119}
\newcommand{\candsCtlHi}{130}
\newcommand{\nCtlImgObj}{102}
\newcommand{\ulFour}{53}
\newcommand{\expFalseFourCons}{1.6}
\newcommand{\nSecOneYear}{12}
\newcommand{\nNoStatic}{42}
\newcommand{\nKeptCtlStrong}{60}
\newcommand{\nKeptCtlStrongSec}{51}
\newcommand{\nKeptCtlInc}{6}
\newcommand{\nSecTwoImg}{123}
\newcommand{\nPassDet}{1,511}
\newcommand{\nStatZero}{505}
\newcommand{\nStatOne}{408}
\newcommand{\nStatTwo}{598}
\newcommand{\pStatTwo}{40}
\newcommand{\nSnrLow}{357}
\newcommand{\nFallback}{60}
\newcommand{\nArchUsed}{1,509}
\newcommand{\resMed}{0.057}
\newcommand{\resNinety}{0.14}
\newcommand{\resMax}{0.37}
\newcommand{\nArchRmsGtTwo}{4}
\newcommand{\nArchRmsGtOneTwo}{68}
\newcommand{\nSecureArcMed}{10.2}
\newcommand{\nProbArcMed}{8.0}
\newcommand{\nCtlWideA}{17}
\newcommand{\nCtlWideB}{12}
\newcommand{\nCtlWideC}{10}
\newcommand{\nCutAlt}{105}
\newcommand{\grpAfterAlt}{182}
\newcommand{\grpBeforeAlt}{1,076}
\newcommand{\nMpsAprFourteen}{139}
\newcommand{\nMpsX}{140}
\newcommand{\nNewNoNight}{1,236}
\newcommand{\nArcObj}{229}
\newcommand{\nMagObj}{216}
\newcommand{\nLongArc}{11}
\newcommand{\longArcLo}{6}
\newcommand{\longArcHi}{15}
\newcommand{\nLongWeightyFour}{7}
\newcommand{\nNoEarlier}{6}
\newcommand{\longRmsLo}{0.03}
\newcommand{\longRmsHi}{0.19}
\newcommand{\longDzMax}{1.1}
\newcommand{\nArcExt}{511}
\newcommand{\nEarlier}{511}
\newcommand{\nDetBefore}{1,432}
\newcommand{\nDetInside}{46}
\newcommand{\nDetAfter}{31}
\newcommand{\nFitTwo}{209}
\newcommand{\nFitThree}{217}
\newcommand{\nFitFour}{91}
\newcommand{\nFitTwoSec}{123}
\newcommand{\nZ}{504}
\newcommand{\zMed}{0.39}
\newcommand{\pZTwo}{1.8}
\newcommand{\zThreeMed}{0.30}
\newcommand{\nZThreeGtThree}{none}
\newcommand{\nZThree}{83}
\newcommand{\expChanceThreeW}{93}
\newcommand{\expChanceFourW}{3.7}
\newcommand{\expFalseThreeUlW}{2.1}
\newcommand{\pCopyDensLo}{0}
\newcommand{\nCopyDensLo}{784}
\newcommand{\pCopyDensHi}{35}
\newcommand{\nMultiStrong}{154}
\newcommand{\nDisjoint}{25}
\newcommand{\nDisjointT}{13}
\newcommand{\nDisjointTSec}{12}
\newcommand{\maxStrong}{30}
\newcommand{\nItfObs}{9,041,468}
\newcommand{\nItfDet}{9}
\newcommand{\nItfObj}{9}
\newcommand{\nVtwoR}{10}
\newcommand{\nVtwoM}{21}
\newcommand{\posLate}{137}
\newcommand{\posDsSeventyEight}{5}
\newcommand{\arcFigN}{517}
\newcommand{\arcFigSec}{396}
\newcommand{\arcFigProb}{121}
\newcommand{\arcFigLong}{11}
\newcommand{\arcFigMedB}{71}
\newcommand{\arcFigMedA}{10.1}
\newcommand{\arcFigGtFive}{456}
\newcommand{\cutObj}{2026~EL$_{143}$}
\newcommand{\cutNcand}{269}
\newcommand{\cutSnrMed}{40}
\newcommand{\cutYears}{2013, 2015, 2017}
\newcommand{\tilestext}{I queried every tile with any part within $20^\circ$ of the ecliptic (barycentric true ecliptic latitude of the cell centre or a corner) and Dec $\leq+35^\circ$, \nTiles\ tiles, in order of ecliptic latitude, with \nReq\ requests and no failed tile. \nTilesRows\ tiles have alerts (Figure~\ref{fig:tiles}), almost all within $10^\circ$ of the ecliptic and at RA $230^\circ$--$360^\circ$, where alerts were issued in 2026; only \nTilesBeta\ lie beyond $10^\circ$. Designated distant objects are found up to $|\beta|=17^\circ$ and none in the few tiles with alerts beyond $20^\circ$, so the band was not widened. The tiles hold \nDia\ DiaSources (one cone row per diaObject plus the sources of multi-source objects), of which \nSel\ pass the selection, on \nNights\ distinct nights between 2026 \firstNight\ and \lastNight. The median tile has alerts on 5 nights. \nDeep\ tiles with more than 15,000 alerts each are deep fields (COSMOS, a field at RA~$186^\circ$ and the Galactic bulge at RA $250^\circ$--$300^\circ$) with up to 38 nights; their links are reported separately.

\begin{figure}[tb]
\centering\includegraphics[width=\textwidth]{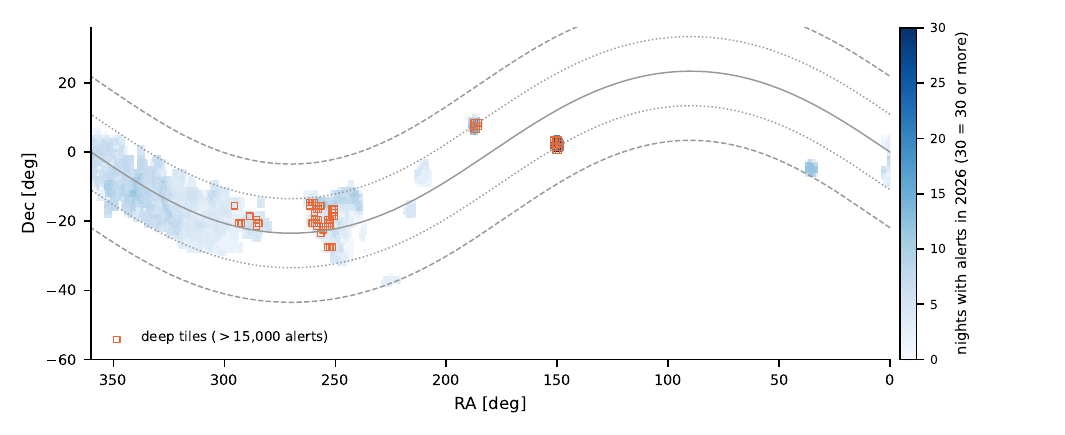}
\caption{The \nTilesRows\ $1^\circ$ tiles with alerts in 2026 (squares at the cell centres, colour = nights with alerts); the deep tiles are outlined. Lines: the ecliptic (solid) and ecliptic latitudes $\pm10^\circ$ (dotted) and $\pm20^\circ$ (dashed).\label{fig:tiles}}
\end{figure}}

\newcommand{\nGrpToJun}{411}
\newcommand{\nGrpJunJul}{560}
\newcommand{\nNewToJun}{764}
\newcommand{\nNewJunJul}{563}
\newcommand{\nNewAfterJul}{289}

\newcommand{\cutTwoObj}{2025~OE$_{697}$}
\newcommand{\cutTwoSnrMed}{5.4}
\newcommand{\cutTwoNcand}{109}
\newcommand{\cutTwoCtl}{2025~MV$_{398}$}
\newcommand{\cutTwoCtlSnr}{67}
\newcommand{\cutTwoCtlN}{20}
\begin{document}

\title{Precovering Rubin Trans-Neptunian Discoveries in Thirteen Years of DECam Archival Data}

\author[0009-0002-5933-3806]{Gioele Cerati}
\affiliation{Independent researcher, Italy}
\email{gioele@cerati.org}

\begin{abstract}
The Vera C.\ Rubin Observatory began discovering distant Solar System objects in 2025, most of them with published arcs of weeks to a few months. Using my software \os, I searched single-night catalogues of 2013--2026 Dark Energy Camera (DECam) archival images for detections of \nSearched\ Rubin discoveries with $a>30$~au or Centaur orbits outside their published observing season. Archival detections were checked on their own images, and for the objects with hits the search was repeated on rotated copies of the published arc as controls. I find such detections for \nImgConf\ objects (\nSecure\ secure, \nProbable\ probable; a few false identifications are expected). They extend the median observed arc from \arcBefore\ to \arcAfter\ and reduce the median uncertainty of the semimajor axis by about two orders of magnitude, and \eIdent\ of the objects are identified with earlier designations. In a secondary search of the 2026 public Rubin alerts within $20^\circ$ of the ecliptic, served by the Fink broker, I recover \nRowsAll\ detections of \nBodies\ known distant bodies that were not associated with them at alert time, \nNew\ of them not among the observations published by the Minor Planet Center by 2026 November~20. Rubin's Prompt Products Database is to carry associations for detections of this kind from its release of 2026 November~30; the recovery here is independent of it and is tested with shifted-sky controls. The published observations of 2025~NJ$_{194}$ and 2000~QK$_{252}$ fit one orbit (rms $\bThreeRms''$), and I identify the two.
\end{abstract}

\keywords{Trans-Neptunian objects (1705) --- Astrometry (80) --- Orbit determination (1175) --- Sky surveys (1464) --- Astronomy software (1855)}

%==========================================================================================
\section{Introduction}\label{sec:intro}

The Vera C. Rubin Observatory \citep{ivezic2019} distributes public alerts for sources detected on difference images \citep{rubinap,rubinppdoi}. In 2026 the alert stream cannot run at full scale, because difference imaging needs templates, which are being built from commissioning and early operations data \citep{rubinearly}. Rubin's prompt Solar System processing compares each detection with the predicted positions of the objects in the Minor Planet Center (MPC) orbit catalogue, and ``the closest DIASource within 1 arcsecond of a known solar system object's predicted position is associated to that object'' \citep{rubinpp}. For the Data Preview~2 reprocessing the documentation adds that the matching did not use positional uncertainties \citep{rubindp2}, and the draft DP2 paper describes it as a one-to-one match to the nearest detection within $1''$ that uses neither brightness nor rate \citep[Section~4.5.1]{rtn115}; for alerts an association is made ``within 1.0 arcseconds of the expected ephemeris position'' \citep{kurlander2026}, and the Rubin Solar System team has confirmed that this description of the $1''$ association applies at alert time (M.~Juri\'c, private communication). A zero \texttt{ssObjectId} means that a detection was not associated at the time of alert generation \citep{kurlander2026}, and in the same thread Kurlander describes an object that Rubin ``rediscovered'' more than $1''$ from its predicted position and whose detections were then submitted. The unassociated detections are linked by HelioLinC3D, Rubin's production implementation of the HelioLinC algorithm \citep{holman2018}, over ``324 different hypotheses spanning heliocentric distances from 1.5 to 9.8 au'' \citep{rubinpp}; the linking page gives tracklets of at least two detections on three nights within 14 days, and other pages give slightly different values \citep{rubinpp}. The DP1 paper adds that ``a dedicated search for objects at heliocentric distances out to 50 au was also conducted'' \citep{rubindp1}. For operations the documentation describes searches for precoveries and isolated detections that can be linked to the refined linkages made by HelioLinC3D \citep{rubinpp}; the radius of these later searches and whether they run today are not documented.

TNOs are found by a separate search, outlined by \citet{holman2026ed138}, with the code \textsc{pointdexter}, which reads Rubin's single-visit direct-image source catalogues and groups detections from different nights. A batch focused on TNOs reported 409 candidates in February 2026 \citep{juric2026forum} and about 380 TNOs were announced in April \citep{rubinnews2026}. The full search cited by \citet{holman2026ed138} found 2,440 previously unknown TNOs in data that end on June~28 (M.~Holman, private communication), that is 2026 June~28, the last night before the formal start of the ten-year Legacy Survey of Space and Time on 2026 June~29, which ended the Early Operations period that followed commissioning \citep[Section~1]{rtn115}, and its catalogue holds objects with at least ten detections on at least six nights within one search region and time window. The LSSTCam data used in their study of the dwarf-planet candidate 2026~ED$_{138}$ run from 2025 May~20 to 2026 July~12, and they extended its arc with DECam precoveries: candidates from the NOIRLab Source Catalog returned by the ADAM precovery service and checked on the images, followed by a measurement of every DECam image along the track, including those after 2019. Rubin reports new discoveries ``on an ad-hoc basis'' \citep{rubinearly}, has sent them in batches, such as one focused on TNOs on 2026 February~16, which also held 2,971 observations of 147 known objects \citep{juric2026forum}, does not submit two-detection tracklets \citep{heinze2026forum}, and states that its searches ``have not been exhaustive'' \citep{juric2026sep}.

Several related efforts use public alerts or archives. An independent project associates Fink alerts with \texttt{ssObjectId}~$=0$ to known orbits with a $10''$ search and joint refits, and reports more than 250,000 added associations and 6,579 orbits with added observations \citep{sibilev2026}; it does not single out distant objects. \citet{hemmann2026} recovered three large-orbit Rubin TNOs of 2025 at their second opposition with CFHT; in two of them the semimajor axis dropped by a factor of about 3--10, and they expected Rubin to ``re-discover'' such objects during the 2026 opposition. The precovery service of the Asteroid Institute's ADAM platform \citep{kiker2023} returns, for an orbit and its uncertainty, the catalogued sources near the predicted positions in the surveys it indexes, which include NSC DR2 (2012--2019) and Rubin detections since 2025 September; it now searches the propagated uncertainty region rather than a fixed $10''$ radius, and its candidates are catalogue matches meant for expert review \citep{adam2026,adamprecovery}. For 2026~ED$_{138}$, 11 of the 28 DECam candidates it returned passed the image check of \citet{holman2026ed138}. THOR, the linker of the same platform, also needs no intra-night tracklets; it was validated on two weeks of ZTF data for $a>1.7$~au, with a single bin of test orbits for 5--50~au \citep{moeyens2021}. Other archives, photographic plates, VISTA, SkyMapper and ATLAS \citep{murtagh2026}, OGLE \citep{mroz2026} and TESS \citep{han2026ed138}, have been searched for 2026~ED$_{138}$ alone. DECam data have been searched for distant objects before: the Dark Energy Survey linked single-night detections without tracklets over its six years \citep{bernardinelli2020,bernardinelli2022}, and the outer Solar System survey of \citet{sheppard2016}, the DECam NEO survey \citep{trilling2017} and DEEP \citep{trilling2024} measured moving objects on many of the exposures used here. The Asteroid Institute has run THOR on the whole of NSC DR2, and its candidates, with arcs mostly shorter than a week, include tens of TNOs \citep{thornsc}. For a known orbit, the CADC's Solar System Object Image Search \citep[SSOS, now SSOIS;][]{gwyn2012} lists the archival images that may contain the object. The present work uses the method of \citet{sibilev2026} for distant objects only, for which a $10''$ radius is too small, and adds a linker that needs one detection per night, with shifted-night controls. It compares what the alerts hold with what the MPC has published, by date. It also searches a decade of DECam data, as \citet{holman2026ed138} did for one object, for all of Rubin's distant discoveries of 2025--2026, in single-night catalogues that for 2020--2026 are my own, with rotated controls and a check of each archival detection on its own image. The linker is \os, my software for linking single-night catalogue detections of distant objects without intra-night tracklets, which was tested on known objects in DECam data in an archival survey \citep{cerati2026survey}; the parts used here are described in Section~\ref{sec:method}.

%==========================================================================================
\section{Data}\label{sec:data}

\subsection{Alerts through the Fink broker}
I used the public REST service of the Fink broker \citep{moller2021,finkapi}, which receives the Rubin alert stream. For each $1^\circ\times1^\circ$ sky tile (a cell of integer RA and Dec, used with a $0\fdg15$ margin) the adapter \cmd{orbitsift.rubin} sends one cone search of radius $0\fdg95$ centred on the cell, with the date window 2026 January~1 to November~21 (which the service applies to the diaObject), which returns the latest alert of each diaObject, and then asks for every source of the diaObjects that have more than one. Queries were made by sky cell in two rate-limited streams. The alert schema versions served were lsst.v10\_0 to v11\_1.

\subsection{Selection}
The adapter keeps DiaSources with positive difference flux, in $grizy$, with none of the flags \texttt{isNegative}, \texttt{glint\_trail} or the bad, edge, saturated, streak, cosmic-ray and no-data pixel flags at the centre, with \texttt{ssObjectId} equal to zero, and belonging to a diaObject seen on one night only. The last condition removes static sources: Rubin groups detections at one position into a diaObject, so a variable star or a bad subtraction returns night after night under one identifier and a mover does not. Fink notes that two alerts less than 6 minutes apart cannot be associated to one diaObject, which happens often in the deep fields \citep{finkissues}; there the condition removes static sources less well. No cut on the real/bogus score (\texttt{reliability}) is made. Magnitudes are $31.4-2.5\log_{10}(f/\mathrm{nJy})$; the mid-exposure time is \texttt{midpointMjdTai} minus 37~s.

Rubin's documentation states that ``each associated DiaSource will either have a non-null diaObjectId or ssObjectId but will not have both fields populated'' \citep{rubinpp}, so the cone search, which is organised by diaObject, should not return associated detections. As a check, the \posN\ associated detections of 2015~BC$_{557}$ that Fink serves under the object's name fall in two of the tiles queried here, and \posIn\ of them are in the cone answers. This is one object in two tiles, so the control is thin; with that caveat, every detection of a known object in this paper is one that Rubin did not associate at alert time.

\subsection{Tiles}
\tilestext

%==========================================================================================
\section{Method}\label{sec:method}

\subsection{Linking}
Within each tile the nights are split into seasons at gaps longer than 45 days, and every season with at least four nights is linked by the single-season linker of \os. Detections are projected on the tangent plane of the tile centre with exact topocentric parallax for Rubin (MPC code X05) and fitted with the parametrisation of \citet{bernstein2000}: pairs of detections on nights at most 40 days apart seed trial orbits at 30--150~au, detections of further nights that fall near the predicted track are added, and a link needs detections on at least four nights, a bound orbit and an rms of at most $0\farcs5$. Intra-night pairs are not required. Only links whose mean position falls in the tile's cell are kept, so that overlapping tiles do not count a link twice.

\subsection{Known-object check}
Each link is fitted with a barycentric two-body orbit and compared with every orbit of the MPC's \texttt{Distant.txt} (\checknorbits\ orbits on 2026 November~20) at the dates of that object's own observed arc; orbits that pass are fitted jointly with the object's published astrometry from the MPC \texttt{get-obs} service (queried by designation only). A link is identified when the joint fit uses at least 90\% of the observations with rms below $0\farcs4$ and none worse than $1\farcs5$. A link that admits no bound two-body orbit cannot be compared and is counted as unidentified.

\subsection{Direct search for known objects}
Objects detected on three nights cannot be linked but can be identified from their catalogued orbits. Independently of the linker, every orbit of \texttt{Distant.txt} was propagated (two-body, X05 topocentre) to each night of each tile, and the detections within $120''$ of the prediction were collected, each compared at its own time. An object is present if detections on at least three nights share a common offset from the prediction within $1\farcs5$: the error of a short-arc orbit over a few weeks is close to a constant shift. A detection is \emph{already reported} if the MPC has an X05 observation of that object within 3~s of its mid-exposure time.

For every designated object I took the first publication of its observations from the references of its MPC observations (a Minor Planet Supplement, or none) and the orbit reference of \texttt{Distant.txt}, and marked each detection as inside or outside the arc of the object's discovery opposition (MPC observations within 200 days of the discovery observation) and as before or after 2026 July~12.

\subsection{Controls}
\os's vetting stage is calibrated on DECam catalogue detections and does not accept Rubin detections, so the controls are the only measurement of chance links here. For each tile-season the linker and the direct search were run again three times on the same detections with every night moved rigidly by a random 20--$60''$ per axis. The shift of a night depends only on the seed and the night, so a night moves the same way in every tile: the controls are three realisations of the nightly shift vectors. No real object can produce a link in a shifted realisation, so the links found there measure the chance rate; the shift is small compared with the separations between unrelated detections that make chance links.

\subsection{Orbit fits}
For a known object with detections that the MPC does not hold, I fitted the MPC observations alone and the MPC observations plus the alert detections with the same code (barycentric two-body fit), with the alert detections at $0\farcs15$. For arcs under 300 days the barycentric distance and radial velocity at the reference epoch were stepped over a grid (first 15--150~au in steps of 1~au, then finer grids over the region of low $\chi^2$), the other four parameters were fitted at each node, and the width $w_a$ is the range of the barycentric semimajor axis over the bound nodes with $\Delta\chi^2\leq2.71$ (scaled by the reduced $\chi^2$ when it exceeds 1), a 90\% profile interval. For longer arcs $w_a$ is the 5--95\% range of $a$ in 300 samples of the linear covariance. The gain is the ratio of the widths before and after; since most arcs are shorter than 300 days before and longer after, it usually compares a profile interval with a covariance interval. For \nWideBefore\ objects, with arcs of \wideArcLo--\wideArcHi~days, the interval before extends beyond $a=1000$~au, towards the limit of bound orbits, so their gain is a lower limit.

\subsection{Archival search for Rubin discoveries}
The archival catalogues are single-night (``orphan'') detections from the NOIRLab Source Catalog DR2 \citep{nidever2021} for 2013--2019 and from my own catalogue of public DECam \citep{flaugher2015} images for 2020--2026, built for the survey of \citet{cerati2026survey}. Rubin discoveries were selected from the MPC's \texttt{Distant.txt} of 2026 November~20: objects with 2025 or 2026 designations and $a>30$~au, or $5<a\leq30$~au with $q>5$~au, whose discovery observation (flag \texttt{disc}) is from X05.

For each object the published observations (all stations, MPC uncertainties, $0\farcs5$ where none is given) were fitted, and the search followed the line of variations (LOV): the coordinate along the long axis of the linear error ellipse of distance and radial velocity, with the other five parameters minimised at each node, out to $\Delta\chi^2=16$ or an unbound orbit. Every catalogue exposure whose $1\fdg15$ footprint the LOV crosses was searched, except within 0.45~d of a published observation, for detections within the across-track uncertainty of the LOV ($\geq1\farcs5$, $\leq60''$) and within 0.8~mag of the predicted magnitude. Candidates were grouped along the LOV, each was refitted with the published observations, and groups were grown greedily; a group is accepted when the orbit is bound, the rms over all observations, published and archival together, is at most $0\farcs2$ and no added detection is worse than $0\farcs6$. A strong hit has detections on at least three nights over at least two seasons other than the published one, or on four such nights, with a spread of magnitude offsets of at most 1.2~mag; these thresholds were fixed before the first target on a validation set. In it, four objects of my survey with multi-year arcs, each started from the 8--25 detections of a single season, were all recovered as strong hits, and 64 false arcs (rotated by $\pm3^\circ$ to $\pm35^\circ$ about the ecliptic pole or shifted in time by 2--4 years) gave 11 hits, all on two nights of one season but one, whose magnitudes differ by 1.4~mag, and no strong hit. As controls, the published arc was rotated about the ecliptic pole by $\pm3^\circ$ and $\pm6^\circ$, with the real residuals added, and searched in the same way; an object is \emph{clean} when no rotated copy gives a strong hit. Objects with more than 10,000 candidates (all in the dense Galactic bulge) were not evaluated.

An archival detection passes the image check when, on its own exposure at the exposure's mid-time, there is a point source with signal-to-noise $\geq5$ and a PSF-like shape within $2''$ of the prediction and $1''$ of the catalogue position, at most 2 of 24 control apertures on the same CCD pass the same test, and no source is found at the same sky position on up to two exposures of other epochs where such exposures exist. A detection measured on a different exposure with the same identifier is flagged. Before/after orbits use the same fitter for both, Find\_Orb \citep{findorb} with DE440 and its default outlier rejection, on the MPC observations alone and with the passing detections at their image positions; the barycentric elements and their uncertainties come from 500 clones of the covariance. The rules for counting an identification are given with the results (Section~\ref{sec:archival}).

%==========================================================================================
\section{Results}\label{sec:results}

\subsection{Archival arcs of Rubin discoveries}\label{sec:archival}
Of the \nPop\ objects in the MPC's \texttt{Distant.txt} with 2025 or 2026 designations and $a>30$~au (\nPopA) or Centaur orbits (\nPopC), the discovery observation is from X05 for \nXfiveAll\ of those with $a>30$~au (\pXfive\%) and \nXfiveC\ of the Centaurs; the MPC observations of all of them were read one by one. Their published arcs are short, weeks to a few months for most. The objects searched below include the \nXfiveC\ Centaurs discovered by Rubin, and \nImgCent\ of the objects of Table~\ref{tab:archival} have Centaur orbits; the others are TNOs.

The archival search was run on \nSearched\ Rubin discoveries; \nCrowded\ in the dense bulge had more than 10,000 candidates and were not evaluated, so \nEval\ were evaluated. \nStrong\ objects have a strong hit, \nWeak\ only weaker hits (two nights in one season, at the level of chance in the validation) and \nNone\ none. For the strong hits the four rotated copies were searched in the same way: \nStrongClean\ objects are clean, \nStrongCtl\ have at least one rotated copy that also gives a strong hit, and for \nStrongInc\ a copy was too crowded to evaluate. The reported hit of an object is its strong result with the most nights; \nMultiStrong\ objects have more than one strong result (up to \maxStrong), and for \nDisjoint\ of them an alternative strong result shares no exposure with the reported one (\nDisjointT\ of these are in Table~\ref{tab:archival}, \nDisjointTSec\ of them secure; they are flagged in the machine-readable table).

\emph{Chance rate.} Of the \nCopies\ rotated copies evaluated, \nCopiesStrong\ (\pCopy\%) give a strong hit; this is the chance rate per search. Chance hits sit at the minimum of the rule: in \nCopiesThree\ of these copies the best strong hit has exactly three archival nights (\pCopyThree\% per search) and in \nCopiesFour\ it has four or more (\pCopyFour\%). Among the real searches, \nKzeroThree\ strong hits have three archival nights and \nKzeroFour\ have four or more. A chance three-night hit becomes the reported hit only where the search has no hit of four or more nights, which is the case for \nNoFour\ of the \nEval\ evaluated objects; about \expChanceThree\ of the \nKzeroThree\ three-night hits are therefore expected to be chance alignments, and about \expChanceFour\ of the hits with four or more nights. The rate grows steeply with the density of candidates along the line of variations, from \pCopyDensLo\% for the \nCopyDensLo\ copies with fewer than 30 candidates to \pCopyDensHi\% for those with 3,000--10,000, and the copies, made only for the strong objects, sample poorer sky than the real searches, which include the weak and empty ones. Weighting the rate in each density bin by the real searches in that bin gives about \expChanceThreeW\ chance three-night hits and \expChanceFourW\ with four or more nights.

\begin{figure}[tb]
\centering\includegraphics[width=0.8\textwidth]{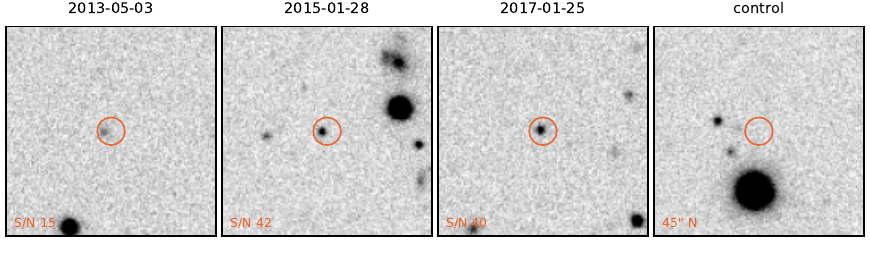}
\caption{Archival DECam detections of \cutObj, the example chosen by the rule of Section~\ref{sec:archival} (median $S/N$ \cutSnrMed): one $30''$ cut per year (\cutYears) from the detection's own exposure, centred on the position predicted by the joint orbit (two-body propagation of the fitted state), with a $2''$ circle on the prediction, the date and the $S/N$ of the image check. Right: the exposure of the last column, centred $45''$ north of the prediction. North is up and east to the left; the grey scale is the same in units of the background noise.\label{fig:cutouts}}
\end{figure}

\emph{Image check.} Each archival detection of a strong hit was measured on its own exposure, one detection on each of up to four seasons of each object (the extreme seasons first); the other detections of each object remain to be checked. The rule is given in Section~\ref{sec:method}. The static-source test used up to two exposures of other epochs where available: of the \nPassDet\ detections that passed for the objects kept below, \nStatTwo\ (\pStatTwo\%) had two such exposures, \nStatOne\ had one and \nStatZero\ had none; \nSnrLow\ have $5\leq S/N<7$ and \nFallback\ were measured on the alternative exposure with the same identifier. Of the detections that failed, \failSnr\ show no source at $S/N\geq5$, \failFrame\ could not be measured on their own exposure (CCD gaps or frames not in the archive), \failCtl\ have more than 2 of 24 control apertures passing, and \failStatic\ have a source at the same position on another epoch. \nConfDone\ of the \nImgDone\ strong objects (\confRate\%) have detections on at least two nights outside the published season that pass. Figure~\ref{fig:cutouts} shows the verified detections of one object, chosen by a fixed rule: of the \cutNcand\ secure objects with verified detections in at least three different years, the one with the highest median $S/N$.

\begin{figure}[tb]
\centering\includegraphics[width=0.6\textwidth]{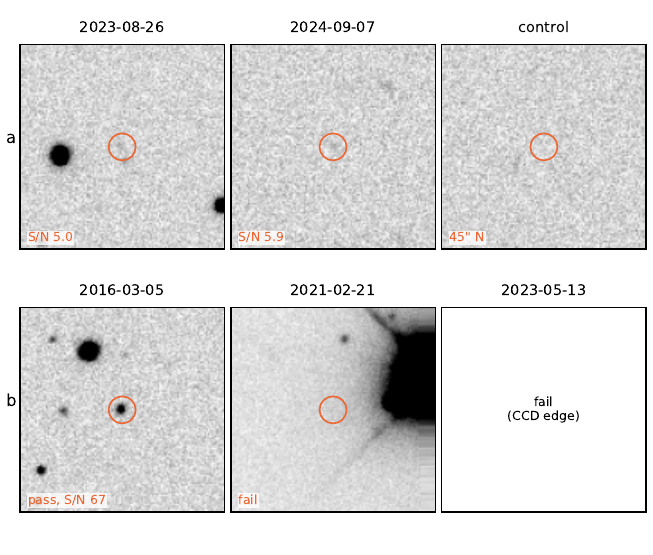}
\caption{The image check at its limits, both examples chosen by fixed rules. (a) \cutTwoObj, the probable object with passing detections in at least two years and the lowest median $S/N$ of its passing detections (\cutTwoSnrMed, among \cutTwoNcand\ such objects): one cut per year, centred on the prediction of the joint orbit, and a control cut $45''$ north. (b) A rotated-arc control copy of \cutTwoCtl, the one whose passing detection has the highest $S/N$ (\cutTwoCtlSnr) of the \cutTwoCtlN\ passing control detections: each checked night of the copy, centred on the catalogue position it linked. The first night is a real point source and passes; the others fail (the 2021 position lies in the halo of a saturated star, and the 2023 one within $15''$ of a CCD edge, so no full cut is shown), so the copy is not counted. Cuts, circle, orientation (north up, east left) and grey scale as in Figure~\ref{fig:cutouts}.\label{fig:cutouts2}}
\end{figure}

The image check shows that a catalogue detection is a real point source on its exposure, not that it is the object. To measure how often a chance alignment survives it, I applied the same check, with the same choice of detections, to the best strong hit of \nCtlImg\ of the rotated copies with a strong hit. Of their \nCtlImgRows\ detections \nCtlImgPass\ pass, and the number of copies with two passing nights outside the published season is \nCtlImgConf, against \confRate\% of the real strong objects: chance hits are made mostly of catalogue detections that are not real point sources. The one-sided 95\% upper limit on the fraction of chance hits that pass is \rcUl\%. Figure~\ref{fig:cutouts2} shows the check at its limits: a probable object with weak passing detections, chosen by a fixed rule, and the strongest passing control detection.

\emph{Identifications.} An object is counted when (i) detections on at least two nights outside the published season pass on their images; (ii) the joint Find\_Orb fit of the MPC observations and the passing detections keeps at least 90\% of all observations with rms at most $0\farcs2$ (\nBadFit\ objects fail this); (iii) that fit keeps detections on at least two archival nights outside the published season, because the rule (ii) is dominated by the published observations and a single archival detection always fits (\nDropBtwo\ objects keep one); and (iv) the catalogue hit is still strong after the nights whose detections failed on images (no source, static source or control apertures) are removed (\nDropBone\ objects become two-night hits). 2025~MR$_{387}$, identified only by a link of an earlier search and with no strong hit in this one, is also left out. These objects are listed as not confirmed in a separate file (Section~\ref{sec:avail}). \nImgConf\ Rubin discoveries remain (Table~\ref{tab:archival}). They are divided into two tiers. \nSecure\ are \emph{secure}: the catalogue hit keeps at least four archival nights and the detections verified on images fall in at least two different years. The other \nProbable\ are \emph{probable}: \nThreeCounted\ of them have three-night hits, and the rest are hits of four or more nights whose verified detections fall in one year (\nSecOneYear) or that kept only three nights after the image check. Of the \nImgConf, \nNoStatic\ have no verified detection with an exposure of another epoch for the static-source test, and \nSecTwoImg\ of the \nSecure\ secure objects have exactly two image-verified nights; the other nights of their hits were not checked. Objects are counted whether or not their rotated controls are clean: \nKeptCtlStrong\ of the \nImgConf\ (\nKeptCtlStrongSec\ of them secure) had a rotated copy that also gave a strong hit, and for \nKeptCtlInc\ a copy was too crowded to evaluate; the column Ctl of Table~\ref{tab:archival} gives this for each object.

\emph{Expected false identifications.} About \expChanceFour\ chance hits of four or more nights are expected before the image check; if such a hit passes the image check no more often than the three-night chance hits measured above, fewer than \expFalseFourUl\ survive, but only \nCopiesFour\ control copies have four or more nights, and on those alone the 95\% limit on the pass rate is \ulFour\%, which would allow up to \expFalseFourCons. For the three-night class, about \expChanceThree\ chance hits are expected before the image check (\expChanceThreeW\ with the density weighting), and with the control measurement at most \expFalseThreeUl\ survive (95\%; \expFalseThreeUlW\ with the weighting); a bound that does not use the expected count is \expFalseThreeCons, the value if every one of the \nThreeNotCounted\ three-night hits that were not counted were a chance alignment. These numbers rest on three assumptions that I cannot test with the present controls: the rotated copies sample sky that is somewhat poorer than the real tracks (a median of \candsCtlLo--\candsCtlHi\ candidates per copy against \candsReal\ for the real searches of the strong objects), the \nCtlImg\ copies come from \nCtlImgObj\ objects, and the limit combines a 95\% bound on the pass rate with a point estimate of the number of chance hits. They count wrong identifications only; a right identification with one wrong detection is not counted, and for the detections used by the fits, each verified on its image and within $0\farcs6$ of an orbit fixed by the others, I expect this to be rare.

Over all observations kept, the \nArchUsed\ archival detections used by the fits have a median residual of $\resMed''$, a 90th percentile of $\resNinety''$ and a maximum of $\resMax''$; the rms of the archival detections alone exceeds $0\farcs12$ for \nArchRmsGtOneTwo\ objects and $0\farcs2$ for \nArchRmsGtTwo. The joint fits use archival detections on two nights outside the published season for \nFitTwo\ objects (\nFitTwoSec\ of them secure), three for \nFitThree\ and four or more for \nFitFour; with two nights the fit absorbs part of the offsets, so for these objects the residuals are a weak test, and the secure ones rest on catalogue nights not checked on images. A test that does not use the residuals is where the joint orbit falls within the uncertainty of the orbit from the published observations alone: chance alignments would be spread over the searched interval of $\Delta\chi^2\leq16$, but for the \nZ\ objects with a bound orbit before, the joint $a$ lies a median of \zMed~$\sigma_a$ from the earlier value, and only \pZTwo\% lie beyond $2\sigma_a$; for the \nZThree\ three-night hits among them the median is \zThreeMed~$\sigma_a$ and \nZThreeGtThree\ lie beyond $3\sigma_a$.

\begin{figure}[tb]
\centering\includegraphics[width=0.48\textwidth]{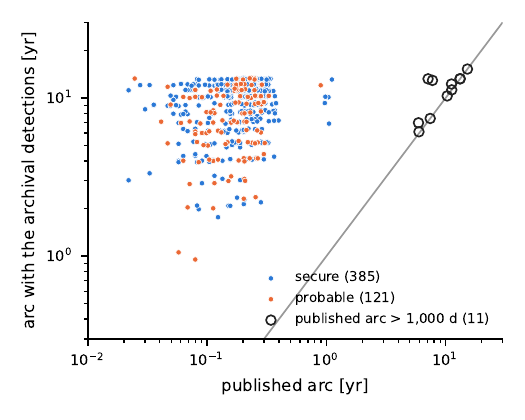}
\caption{Published arc against the arc with the archival detections used by the joint fits, for the \arcFigN\ objects of Table~\ref{tab:archival} (\arcFigSec\ secure, \arcFigProb\ probable). Circles: the \arcFigLong\ objects, all secure, whose published arc was already longer than 1,000 days (the test with a known answer). Solid line: no change. The median arc grows from \arcFigMedB~days to \arcFigMedA~years, and \arcFigGtFive\ objects reach more than five years.\label{fig:arcs}}
\end{figure}

\begin{figure}[tb]
\centering\includegraphics[width=\textwidth]{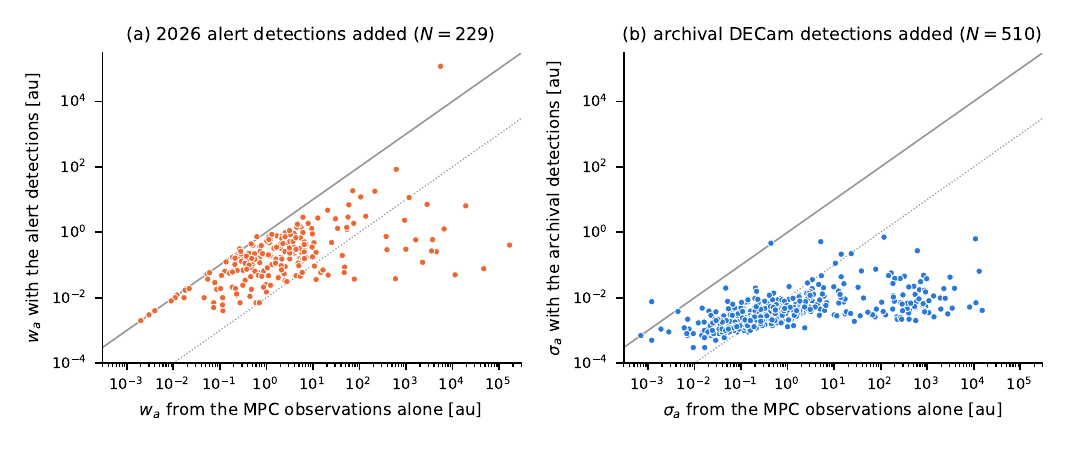}
\caption{(a) Width $w_a$ of the 90\% interval of the barycentric $a$ (two-body fits; Section~\ref{sec:method}) from the MPC observations alone and with the unassociated 2026 alert detections added. (b) $\sigma_a$ from Find\_Orb fits with the same settings to the MPC observations alone and with the archival DECam detections, for the objects of Table~\ref{tab:archival} with a value of $\sigma_a$ in both fits ($N$ in the panel title). Solid line: no change; dotted: a factor of 100.\label{fig:gain}}
\end{figure}

\emph{Orbits.} Figure~\ref{fig:arcs} compares the arcs before and after. Of the archival detections used by the fits, \nDetBefore\ are earlier than the first published observation of their object, \nDetInside\ fall inside the published arc and \nDetAfter\ after it; \nEarlier\ objects have detections earlier than their published arc. \nLongArc\ objects already had published arcs of \longArcLo--\longArcHi~years (\nLongWeightyFour\ of them with DECam observations at the MPC), and for \nNoEarlier\ of them every archival detection falls inside the published arc, which they do not extend. These \nLongArc\ objects are a test with a known answer: their joint fits keep every archival detection, with an archival rms of \longRmsLo--$\longRmsHi''$, and the joint $a$ lies within \longDzMax~$\sigma_a$ of the value from the published observations alone. For the \nFits\ objects, the archival detections used by the fits extend the median arc from \arcBefore\ to \arcAfter\ (\nSecureArcMed~years for the secure and \nProbArcMed~years for the probable tier), and $\sigma_a$ from the same fitter falls by a median factor of \sigGainMed\ to a median of \sigAfterMed~au (Figure~\ref{fig:gain}b). The $\sigma_a$ before comes from the linear covariance of arcs of weeks to months; against the 90\% interval of the scan of Section~\ref{sec:method} on the same observations (two-body), it is too small by more than a factor of ten for \aCmpWide\ objects, all with arcs of at most \aCmpArcHi~days, and with $w_a/3.29$ of the scan in place of $\sigma_a$ before the median factor is \aCmpGain\ (\aCmpN\ objects). With the same fitter before and after, the MPC observations alone give no bound orbit for \nUnbBefore\ objects, and $a$ differs by more than 10\% of the joint-fit value for \nAoffLike\ of the other \nBoundBefore\ (\nAoffSel\ of them among the \nSelTargets\ objects whose published $a$ exceeded 150~au); against the MPC catalogue $a$ the count is \nAoffCat, \nAoffCatU\ of them with $U=8$ or 9. These are one-opposition orbits that the MPC marks as uncertain, so the changes are expected rather than errors of the MPC orbits; they confirm on a larger sample what \citet{hemmann2026} found for two of their three objects.

\emph{Earlier designations.} A Rubin discovery may be an object designated earlier, as 2025~NJ$_{194}$ is (Section~\ref{sec:nj194}). For each of the \eScreenOld\ objects of \texttt{Distant.txt} designated before 2025, I propagated its MPC orbit and the joint orbit of every object of Table~\ref{tab:archival} to the date of its last observation; \eScreenCand\ pairs agree within $\eScreenTol^\circ$ and 2~mag in $H$. Scaled from the pairs at $0\fdg3$--$1^\circ$, \eExpTenth\ pairs are expected by chance within $0\fdg1$ and \eExpHundredth\ within $0\fdg01$; \eWithinHundredth\ pairs lie within $0\fdg01$ (all within $10''$) and \eChanceTenth\ more within $0\fdg1$. For these pairs, for three pairs found by a comparison of elements, and as controls for \eCtlN\ random pairs at $0\fdg1$--$0\fdg5$, I fitted the published observations of both designations and the archival detections jointly with Find\_Orb. I count a pair as identified when the fit keeps at least 90\% of all observations, at least one observation on every night of the earlier designation (at least two nights), and has rms at most $0\farcs25$, looser than the $0\farcs2$ of Table~\ref{tab:archival} because the older astrometry has errors of 0.3--$0\farcs5$; as possible when the rms condition holds and some but not all nights of the earlier designation are kept. \eIdent\ pairs are identified: \eIdentList. \ePossible\ are possible (\ePossList): in each, the fit keeps the first night or nights of the earlier designation and rejects a later night as a whole. The pairs within $0\fdg1$ that are not within $0\fdg01$, the three element-based pairs and the \eCtlN\ controls do not fit (\eCtlIdent\ identified and \eCtlPossible\ possible among the controls). For \ePredN\ of these pairs and \ePredCtl\ of the controls Find\_Orb did not converge from any of the three starts; for them the orbit of Table~\ref{tab:archival}, propagated (two-body) to the dates of the earlier designation, misses its observations by $\ePredMin'$ to $\ePredMax^\circ$, whereas for 2026~GT$_{27}$ and 2003~GM$_{53}$ it passes within $\ePredPos''$ of all of them.

Three of the identified pairs need a remark. The nine MPC positions of 2016~FH$_{59}$ repeat the same position within each night while the object moves about $1''$ in 20 minutes, which explains the rms of the joint fit with 2026~EP$_{148}$; one archival detection of 2026~EP$_{148}$ is an exposure already at the MPC as an observation of 2016~FH$_{59}$ and is left out of the fit and of the files below. The eight observations of 2022~GN$_{14}$ (station W93) are offset by about $0\farcs5$ in the same direction from the orbit anchored by the archival detections of 2021 and 2023, and the fit of the published observations alone does not converge in Find\_Orb. 2025~NK$_{497}$ fits 2002~PE$_{153}$ only when the fit starts from the orbit of Table~\ref{tab:archival}; 2002~PD$_{153}$, measured on the same exposures $3\farcm7$ away, fails, which is a built-in control. For the possible pair 2025~ML$_{352}$ and 2004~MM$_{10}$ the fit that keeps the observations of 2004 rejects two of the three archival detections of 2025~ML$_{352}$, so either the pairing or those detections are wrong; the object is flagged in the machine-readable table. With 2025~NJ$_{194}$ = 2000~QK$_{252}$ this makes \eIdentAll\ identifications in all. The MPC identifier service lists none of these designations as linked (2026 November~24); I identify the pairs pending confirmation by the MPC and have not reported them. The fits are a supplement, and the identifications of Table~\ref{tab:archival} are not a claim on discovery credit.

\emph{Positions.} The \posRows\ archival detections of the \posObj\ objects that the fits used, with the exposure, filter, signal-to-noise and a magnitude on the Gaia $G$ scale, are given as a plain table (Section~\ref{sec:avail}); \posInSeason\ of them lie within 200 days of a published observation, \posSnrLtEight\ have $S/N<8$ and \posSnrLtSeven\ have $S/N<7$, and \posDropTime\ detections whose exposure header was not available locally, so that their time could not be checked, are left out; with these and the duplicate row of 2026~EP$_{148}$, \posOneRow\ objects (\posOneRowList) keep a single position in the table, although their fits used two archival nights. \citet{holman2026ed138} required $S/N\geq8$ for their DECam precoveries; with that cut and two nights outside the published season, \adesEightObj\ objects (\adesEightSec\ of them secure) keep \adesEightRows\ detections. Every image was retrieved from the Astro Data Archive without credentials, which returns only data out of their proprietary period; \posLate\ positions are from exposures of 2025 and 2026. \posDsSeventyEight\ detections of 2026~DS$_{78}$, from a second pass over its images, lie within 0.45~d of a published observation, on other exposures. In the MPC's Isolated Tracklet File of 2026 November~22 (\nItfObs\ observations) \nItfDet\ of the detections used by the fits, of \nItfObj\ objects, are already present as unlinked observations from W84 (same time within 2~minutes and position within $3''$); the identifications link them to their objects. \nVtwoR\ objects of Table~\ref{tab:archival} were already identified, with DECam positions, in \citet{cerati2026survey}, and \nVtwoM\ more were listed there as candidate matches to designated objects. None of the detections has been submitted to the MPC. All \posRows\ detections are given in ADES format with the paper (Section~\ref{sec:avail}), for submission through the MPC's route for archival data after its review; \adesEightRows\ of them, of \adesEightObj\ objects, have $S/N\geq8$ on two nights outside the published season.

\subsection{Unassociated detections of designated distant objects}\label{sec:known}
The detections of this section were not associated at alert time. The Prompt Products Database (PPDB), scheduled for release on 2026 November~30 \citep{guy2026ppdb}, is to carry associations for such detections, and the PPDB is to update its table of potential associations, NearbySSO, daily with the latest MPC orbit catalogue, as confirmed by the Rubin Solar System team (M.~Juri\'c, private communication). The PPDB was not available for this paper and is not used in the counts below. This version was written before the PPDB release; version 2 of this paper will compare these detections with the PPDB associations and give how many of the \nNew\ it holds. What this section adds to the associations that Rubin is to publish is (i) a recovery of the detections that does not depend on them, with the chance rate measured on shifted-sky controls (Section~\ref{sec:controls}); (ii) their status at the MPC on 2026 November~20; and (iii) the identification of 2025~NJ$_{194}$ with 2000~QK$_{252}$ (Section~\ref{sec:nj194}), which, with the \eIdent\ identifications of Rubin discoveries with earlier designations from the archival search (Section~\ref{sec:archival}), does not follow from an association with a single catalogued orbit.

\begin{figure}[tb]
\centering\includegraphics[width=\textwidth]{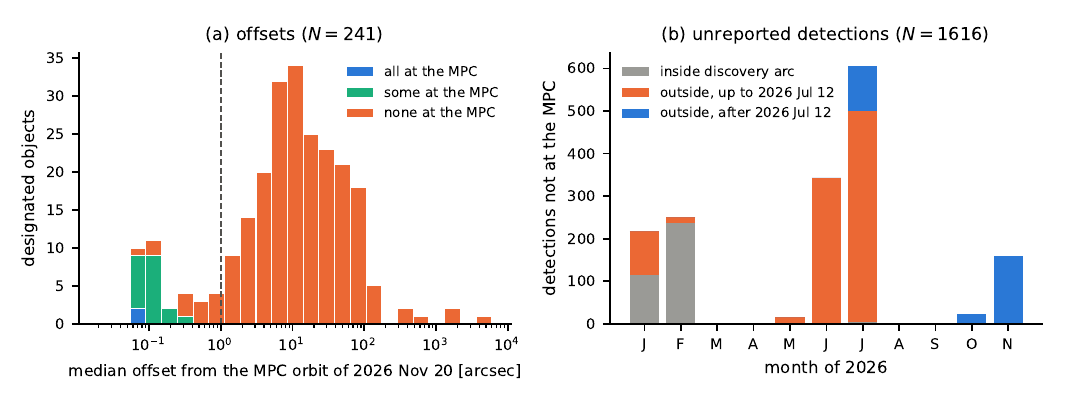}
\caption{(a) Median offset of the unassociated alert detections of the \nObj\ designations from the MPC orbit of 2026 November~20, coloured by whether the MPC holds the detections (an X05 observation within 3~s); the dashed line is Rubin's $1''$ association radius, which applies only to objects catalogued at alert time. Objects whose detections are at the MPC have small offsets because the orbit was fitted to them. (b) The detections not at the MPC by month of 2026: inside the object's discovery-opposition arc, and outside it before and after 2026 July~12. The gaps follow the months in which Fink served alerts in these tiles.\label{fig:offsets}}
\end{figure}

The direct search finds \nObjOffset\ designated objects in the real detections, and none in any of the three shifted realisations; the linker adds \nObjLinkOnly\ more that the direct search missed, most of them beyond its $120''$ radius. These \nObj\ designations belong to \nBodies\ bodies, because 2025~NJ$_{194}$ and 2000~QK$_{252}$ are the same object (Section~\ref{sec:nj194}). \nObjXfive\ of the designations (\nBodiesXfive\ of the bodies) are Rubin discoveries, nearly all of 2025 with one opposition and published arcs of weeks to months, detected again in 2026. For the \nMagObj\ objects with new detections outside their discovery arc, the median magnitude of an object's detections is \magMed\ in the median (from \magLo\ to \magHi). The median offset of the \nRowsAll\ distinct unassociated detections from the MPC orbits of 2026 November~20 is $\offMedDet''$ ($\offMedNew''$ for the \nNew\ not at the MPC), and the median object has a median offset of $\offMed''$ (Figure~\ref{fig:offsets}a); of the \nTab\ designations of Table~\ref{tab:alerts}, \offGtOne\ are beyond $1''$ and \offGtFive\ beyond $5''$. The offsets lie mostly along the ecliptic, as expected for orbits from one opposition. For objects with small offsets the current MPC orbit already includes Rubin observations of the same nights, so a small offset now says nothing about the offset at alert time.

As of 2026 November~20, \nNew\ of the detections, on \nNewNights\ object-nights, have no X05 observation of the same object within 3~s among the MPC's published observations, and for \nNoneAtMPCbody\ bodies the MPC holds none of them (Table~\ref{tab:alerts}). \nNewNoNight\ of them are on nights for which the MPC holds no observation of the object; the others fall on nights for which it holds one. By date they fall into three exclusive groups (Figure~\ref{fig:offsets}b). \nGrpArc\ lie inside the object's own discovery-opposition arc, most of them in the deep fields, where Rubin discovered the objects in the same data. \nGrpAfter\ lie outside it and date from after 2026 July~12, the last date of the LSSTCam data described by \citet{holman2026ed138}. The remaining \nGrpBefore\ lie outside the discovery arc and before that date: \nBefJunJul\ are from June and July 2026 and \nBefEarly\ from January, February and May; \nBefXtwentyfive\ belong to Rubin discoveries of 2025, \nBefXtwentysix\ to Rubin discoveries of 2026 seen outside the 200-day window of their arc, and \nBefOther\ to objects discovered in 2000--2023 at other stations. Split at the end of the data of Rubin's TNO search (the night of 2026 June~28; Section~\ref{sec:intro}), \nGrpToJun\ of these \nGrpBefore\ are from nights up to June~28, within the data of that search, and \nGrpJunJul\ from June~29 to July~12, after them. Over all \nNew\ new detections, inside the discovery arcs included, the three periods (up to June~28, June~29 to July~12, after July~12) hold \nNewToJun, \nNewJunJul\ and \nNewAfterJul. Of the \nGrpBefore, \nBefRubinTwentyfive\ are June and July detections of Rubin's 2025 discoveries, the second-opposition detections that \citet{hemmann2026} expected Rubin to ``re-discover'' during the 2026 opposition. The cut at July~12 is the last date of the data set of \citet{holman2026ed138}; the MPC's latest X05 observation of any of these objects is of 2026 \latestXfive, and over all of Rubin's distant discoveries of July~14. \nCutAlt\ of the \nGrpAfter\ detections after the cut are from the two following nights (UTC July~13 and 14); with the cut at July~14 the three groups are \nGrpArc, \grpAfterAlt\ and \grpBeforeAlt. Rubin states that its searches ``have not been exhaustive'' and remain in a ``best effort'' mode \citep{juric2026sep}; it submitted observations of known objects in large batches (698,926 observations of 64,346 known objects on 2026 February~2, \citealt{juric2026forum}), and not all measurements of known objects have yet been reported to the MPC, which it does ``in stages'' \citep{rubindp2}. The numbers measure how far the public alerts run ahead of the reports on this date, not a failure of Rubin's processing.

\emph{Publication history.} The publication history of the objects separates two classes. (i) \nMpsObj\ bodies (\nMpsObjX\ Rubin discoveries of 2025 and \nMpsObjOther\ objects discovered elsewhere before 2026) have observations first published in a Minor Planet Supplement. For \nMpsAprFourteen\ of the \nMpsX\ Rubin objects the supplement that first published them belongs to the batch of 2026 April~14 \citep{mpcarchive}, which holds observations up to 2026 \mpsLbDate; on 2026 April~2 Rubin had announced about 380 TNOs whose data, it wrote, had been confirmed by the MPC \citep{rubinnews2026}. These objects were therefore designated by 2026 April~14 at the latest. They hold \nNewMps\ of the \nNew\ new detections (\nMpsBefore\ before and \nMpsAfter\ after July~12, none in a discovery arc). \nMpsLate\ of them (\pMpsShare\% of all new detections, \nMpsLateObj\ bodies) were taken after 2026 April~15; their median offset from the current MPC orbit is $\mpsLateOff''$ and \pMpsLateGtOne\% lie more than $1''$ from it. I infer that these objects were probably catalogued and that most of these detections missed the $1''$ radius because one-opposition orbits are off by several arcseconds at the next opposition, the situation that \citet{kurlander2026} describes and that \citet{hemmann2026} anticipated. Two caveats apply. Designated is not the same as present in the orbit catalogue that prompt processing used on the night, whose refresh cadence is not documented in the sources cited; only \mpsLateCat\ of the \nMpsLate\ detections are later than the orbit reference of the current catalogue entry. And the offsets are from the orbits of 2026 November~20, whose references for these objects date from \mpsRefLo\ to \mpsRefHi\ 2026, after most of the detections. \mpsLateLeOne\ of the \nMpsLate\ lie within $1''$ of the current orbit and were still not associated; either the orbit has changed since, or the object was not in the catalogue used. (ii) \nNorefObj\ designations have no publication reference on any observation and an orbit only in MPECs of late July to October 2026. They hold \nNewNoref\ new detections: \nNorefArc\ inside the discovery arc, \nNorefBefore\ before and \nNorefAfter\ after July~12. Most of these detections predate the MPEC, and for them ``not yet designated when the alert was issued'' is the likely reason. In addition, the Fink documentation notes that before 2026 February~20 Rubin associated no alert with a known asteroid \citep{finkissues}; \nNewEarlyFeb\ of the new detections are earlier than that date. The designation dates are bounded by the date of the supplement batch for class (i) and by the MPECs for class (ii); the Rubin team has the exact dates. A stricter test, an orbit reference of the MPC catalogue older than every alert detection of the object, holds for \nCatAll\ of the \nObj\ designations; it uses the latest orbit reference, which is rewritten at every orbit update, so it can show that an object was catalogued but not that it was not.

\subsection{Orbit improvement}\label{sec:gain}
For the \nTab\ designations with new detections and a fit (Table~\ref{tab:alerts}), the joint fit of the MPC observations and the alert detections has an rms of at most $\rmsMax$, and no alert detection has a residual larger than $\oursMax''$; the fits are consistent with the identifications. Leaving out the \nArcOnly\ objects whose new detections all lie inside their own discovery arc, the detections extend the median arc of these \nArcObj\ objects from \arcBeforeAl\ to \arcAfterAl~days, and the width $w_a$ of the 90\% interval of $a$ narrows from a median of \wBefore\ to \wAfter~au, by a median factor of \gainMed\ (more than 10 for \gainGtTen\ and more than 100 for \gainGtHundred\ of \nGainObj\ objects; less than 1.5 for \gainLtOneFive, and \gainLtOne\ get wider; Figure~\ref{fig:gain}a). For the few objects with published arcs of more than a decade the two-body fit is only a test of the identity. The ``before'' widths are from my own fits to the MPC observations; against the MPC catalogue, the joint $a$ differs by more than 10\% for \nBigA\ objects, all of which have uncertainty parameter $U=8$ or 9, orbits that the MPC itself marks as poorly determined, where a change is expected.

\begin{deluxetable*}{lcclrcrcrrrrr}
\tabletypesize{\scriptsize}\tablewidth{0pt}
\tablecaption{Designated distant objects with unassociated 2026 alert detections not reported to the MPC as of 2026 November~20 (first 10 of 237 rows)\label{tab:alerts}}
\tablehead{\colhead{Object} & \colhead{Disc.} & \colhead{Pub.} & \colhead{$N_\mathrm{MPC}$} & \colhead{Arc} & \colhead{New} & \colhead{Offset} & \colhead{L} & \colhead{$a_\mathrm{MPC}$ (U)} & \colhead{$a$} & \colhead{$w_a$ before} & \colhead{$w_a$ after} & \colhead{rms}\\
\colhead{} & \colhead{} & \colhead{} & \colhead{} & \colhead{(d)} & \colhead{all/pre/arc} & \colhead{($''$)} & \colhead{} & \colhead{(au)} & \colhead{(au)} & \colhead{(au)} & \colhead{(au)} & \colhead{($''$)}}
\startdata
2000~QK$_{252}$ & 807 2000-08 & MPS & 4 & 27 & 4/4/0 & 5602.5 & y & 45.15 (E) & 46.91 & 46826 & 0.077 & 0.13 \\
2003~QY$_{113}$ & 568 2003-08 & MPS & 27 & 8106 & 3/0/0 & 3.5 &  & 44.87 (5) & 44.87 & 0.074 & 0.007 & 0.17 \\
2006~OC$_{22}$ & 568 2006-07 & MPS & 45 & 6999 & 4/4/0 & 0.7 & y & 43.15 (7) & 43.27 & 0.077 & 0.005 & 0.14 \\
2013~EV$_{134}$ & 695 2013-03 & MPS & 17 & 4840 & 1/1/0 & 0.1 & y & 39.39 (5) & 39.50 & 0.012 & 0.012 & 0.12 \\
2013~NG$_{83}$ & W84 2013-07 & MPS & 24 & 4394 & 3/3/0 & 1.8 &  & 39.31 (6) & 39.41 & 0.477 & 0.018 & 0.09 \\
2014~WF$_{626}$ & T09 2014-11 & -- & 116 & 4154 & 13/13/0 & 0.1 &  & 45.56 (5) & 45.26 & 0.053 & 0.053 & 0.15 \\
2015~BO$_{641}$ & T09 2015-01 & MPS & 424 & 4059 & 14/14/0 & 0.1 & y & 46.15 (4) & 45.86 & 0.017 & 0.010 & 0.15 \\
2015~FJ$_{561}$ & T09 2015-03 & MPS & 46 & 4171 & 4/4/0 & 0.4 &  & 41.97 (5) & 42.11 & 0.011 & 0.010 & 0.17 \\
2015~KY$_{162}$ & W84 2015-05 & MPS & 26 & 3765 & 5/5/0 & 3.6 & y & 38.34 (5) & 38.43 & 0.046 & 0.010 & 0.15 \\
2015~TQ$_{454}$ & W84 2015-10 & MPS & 7 & 796 & 6/0/0 & 28.4 & y & 51.70 (5) & 51.51 & 0.104 & 0.049 & 0.14 \\
\enddata
\tablecomments{Table~\ref{tab:alerts} is published in its entirety in machine-readable form (\texttt{table2\_alerts.csv}); a portion is shown here for guidance regarding its form and content. Disc.: discovery station and month. Pub.: first publication of the object's observations (MPS: in a Minor Planet Supplement, for 139 of the 140 Rubin objects the batch of 2026 April~14; --: no publication reference on any observation, orbit only in an MPEC). $N_\mathrm{MPC}$, Arc: published observations and arc. New: unassociated alert detections with no X05 observation at the MPC within 3~s: all, those dated up to 2026 July~12, and those inside the object's discovery-opposition arc. Offset: median offset of the detections from the MPC orbit of 2026 November~20. L: also linked by the linker and identified. $a_\mathrm{MPC}$ (U): heliocentric semimajor axis and uncertainty parameter of that orbit. $a$: barycentric, joint two-body fit of the MPC observations and the alert detections. $w_a$: width of the 90\% interval of the barycentric $a$ (Section~\ref{sec:method}), from the MPC observations alone and with the alert detections. rms: joint fit. The table has 237 rows for 241 designations: the others have no new detection or no joint fit. I identify 2000~QK$_{252}$ with 2025~NJ$_{194}$ (Section~\ref{sec:nj194}); three detections are common to their two rows.}
\end{deluxetable*}

\subsection{Linking, identifications and controls}\label{sec:controls}
In the wide-survey tiles the real detections give \nLinksWide\ links in tile cores; \nLinksWideId\ are identified with designated objects, with rms up to $\rmsIdHi''$, spanning \idDaysLo--\idDaysHi~days. The three shifted realisations give \nCtlWideTot\ links, \nCtlWide\ per realisation, all on four nights, at rms \rmsCtlLo--$\rmsCtlHi''$ and spanning \ctlDaysLo--\ctlDaysHi~days; \nCtlWideNoorb\ of them admit no bound orbit, and \ctlTopTile\ come from a single tile. Of the \nLinksWideUn\ unidentified real links, \nLinksWideUnFour\ have four nights, a number consistent with the chance rate (for a Poisson mean of \nCtlWide, $P(\geq\nLinksWideUnFour)=0.32$ and $P(\geq\nLinksWideUn)=\pGe$; the controls are more clustered than a Poisson process, \nCtlWideA, \nCtlWideB\ and \nCtlWideC\ per realisation, so these values are indicative; \nLinksWideUnNoorb\ admit no bound orbit). Their rms values are not: \unidFourRest\ resemble the control links (rms \unidFourRestRmsLo--$\unidFourRestRmsHi''$, \unidFourRestDaysLo--\unidFourRestDaysHi~days), whereas \unidFourLow\ have rms \unidFourLowRmsLo--$\unidFourLowRmsHi''$, below that of every control link, and span \unidFourLowDaysLo--\unidFourLowDaysHi~days like the identified links; these are probably real undesignated objects. The other three unidentified links have five, six and ten nights, where the controls give none, and rms below $0\farcs1$; all three also match undesignated detections in archival DECam data and are probably real. A second check of the \nLinksWideUn\ unidentified links samples no orbits: every orbit of the MPC's \texttt{Distant.txt} of 2026 November~24 and of \texttt{MPCORB.DAT} of November~23 with $a>5.5$~au was propagated to the dates of the link, and the link's orbit, on a grid of distances of 20--100~au (5--300~au when no bound orbit exists in that range; for \uNoBound\ links none exists, and only the first step applies), was propagated to the dates of each catalogued object's own arc. The \uPairs\ object--link pairs within $2^\circ$ or $3^\circ$ of each other were fitted locally with five positions of the object's MPC orbit, and the \uFits\ pairs with an rms below $2''$ were fitted jointly with the object's published astrometry. None matched; one joint Find\_Orb fit failed, and for that pair, an object with a one-day arc, the best two-body fit from 285 starting orbits is unbound with an rms of $1\farcs9$. The same check identifies all \uCtlN\ identified links with the shortest arcs, used as a test. No position of an unidentified link is given. In the wide-survey tiles the linker links \recallN\ of the \recallOf\ objects that the direct search sees on at least four nights (95\% interval \recallLo--\recallHi\%).

In the \nDeep\ deep tiles the linker gives \nDeepReal\ links, \nDeepRealFour\ of them on four nights, against \nDeepCtl\ per shifted realisation: the four-night links there are consistent with chance. \nDeepId\ links are identified, and \nDeepUnFive\ unidentified links with five or more nights compare with \nDeepCtlFive\ in the three realisations together (\deepCtlFivePer\ expected per realisation). The linker links \recallDeepN\ of the \recallDeepOf\ objects that the direct search sees on four nights in these tiles. No result of this paper depends on the deep tiles' unidentified links.

\begin{figure}[tb]
\centering\includegraphics[width=\textwidth]{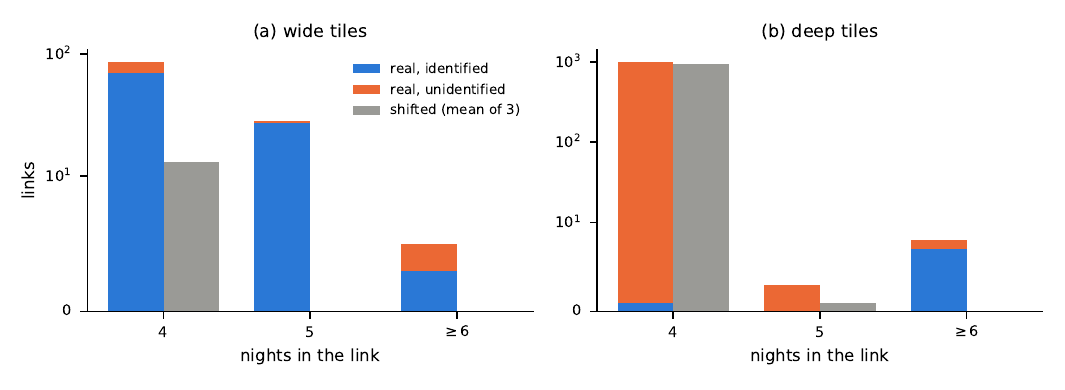}
\caption{Links in tile cores by number of nights (logarithmic scale above 10): links of the real detections, identified with a designated object or not, and the mean of three realisations with each night shifted rigidly by 20--$60''$. (a) Wide-survey tiles; (b) the \nDeep\ deep tiles.\label{fig:controls}}
\end{figure}

\subsection{The identity of 2025~NJ$_{194}$ and 2000~QK$_{252}$}\label{sec:nj194}
The linker found a link that the known-object check matched to two designations: 2000~QK$_{252}$, a one-opposition object with \bThreeOld\ observations over \qkArc~days at station 807 and a prediction $\qkOff^\circ$ from the detections, and 2025~NJ$_{194}$, a Rubin discovery of 2025 with \bThreeNew\ observations. The direct search had found three of the same detections under 2025~NJ$_{194}$. A Find\_Orb fit of the published observations of both designations alone keeps \bThreeUsed\ of \bThreeCount\ observations with rms $\bThreeRms''$ (largest residual $\bThreeMax''$) and gives a barycentric $a=\bThreeA\pm\bThreeAsig$~au, $e=\bThreeE$ and $i=\bThreeI^\circ$; with Find\_Orb's outlier rejection switched off it keeps \bNorejUsed\ of \bNorejCount\ at rms $\bNorejRms''$. As a prediction test, the orbit fitted to the 2025 observations and the \bNAlert\ alert detections of 2026 alone (rms $\bPredRms''$, $a=\bPredA\pm\bPredAsig$~au) places the four positions of 2000 \bPredOffLo--\bPredOffHi$'$ from its prediction, whose one-sigma uncertainty along the line of variations is about \bPredSd$'$. Across the line of variations the four positions lie within $\bPredCross''$ of the predicted strip, whose one-sigma width is about $\bPredCrossSd''$; along it they lie at \bPredAlongSig~sigma of a strip about \bPredStrip$'$ long. The test is consistent but only fails to refute; the identification rests on the joint fit. I identify 2025~NJ$_{194}$ with 2000~QK$_{252}$, pending confirmation by the MPC; if the MPC accepts it, the 2025 observations become a recovery of the object of 2000, as for 2025~MX$_{348}$, which the MPC linked to 2015~KF$_{176}$ \citep{hemmann2026}. The MPC identifier service does not list either designation as linked to the other (2026 November~24), and I have not reported the identification. Four other links also matched two designations (2025~MV$_{399}$ and 2002~GT$_{32}$, 2025~NQ$_{193}$ and 2002~PN$_{147}$, 2025~OE$_{343}$ and 2002~PN$_{153}$, 2025~OU$_{344}$ and 2002~PF$_{149}$); the arcs of 2002 are one to two days long, and these four objects are not in Table~\ref{tab:archival}, so no decade-long archival arc is available to test the identity as in Section~\ref{sec:archival}.

%==========================================================================================
\section{Summary}\label{sec:summary}

The main result of this paper is a set of archival arcs for Rubin's distant discoveries. Archival DECam detections outside the published observing season, verified on their own images, were found for \posObj\ Rubin discoveries (\nSecure\ secure, \nProbable\ probable); for \nArcExt\ of them they extend the observed arc, and the median arc grows from \arcBefore\ to \arcAfter. The detections are given as a table, and as an ADES file (measurer G.~Cerati), for submission through the MPC's archival route. The identity 2025~NJ$_{194}$ = 2000~QK$_{252}$, the \eIdent\ identifications of Rubin discoveries with earlier designations in Section~\ref{sec:archival} and the \ePossible\ possible ones have not been reported and will be offered to the MPC for confirmation.

The alert search is a secondary result. Its \nRowsAll\ unassociated detections of \nBodies\ known bodies, \nNew\ of them not at the MPC on 2026 November~20, are of the kind that Rubin's Prompt Products Database is to associate from its release of 2026 November~30, with its table of potential associations refreshed daily from the latest MPC orbit catalogue (version 2 of this paper will compare them with that release). What remains specific to this paper is an independent recovery of these detections with shifted-sky controls, their status at the MPC on 2026 November~20, and the identifications.

%==========================================================================================
\section{Limitations}\label{sec:limits}

Links of Rubin detections are not vetted: the vetoes of \os\ are calibrated on DECam detections. A link is believed here only when it is identified with a designated object, and unidentified links are compared with the shifted controls and nothing more is claimed for them; their alert cutouts were not examined. No conclusion rests on an object not being detected. The known-object check, the direct search and the orbit fits of the alert detections use two-body orbits, adequate over arcs of weeks to months and only an identity test over decades; the archival fits and the identity tests use Find\_Orb with planetary perturbations.

The counts apply to the 2026 alert stream as served by one broker (Fink), where alerts were issued; detections from before the start of alert production in February 2026 \citep{rubinearly} are those that Fink served, and I did not check against Rubin's own archive whether the cone search returns every alert in the date window. The \nBodies\ bodies are a lower limit with no completeness estimate: the direct search needs three nights within $120''$ of the prediction and the linker four nights. The median tile has \densMedNight\ selected detections per square degree per night, far below the LSST design density; the linker's chance-link rate at the design density is unmeasured, and the REST interface is adequate for a region but not for the whole sky (\nReq\ requests and \fetchHours~hours here). The offsets are measured from the MPC orbits of 2026 November~20, which already include observations made after the alerts. The designation dates behind the two classes of Section~\ref{sec:known} are upper bounds, from the date of the supplement batch and of the MPECs.

The archival search covers only the catalogues named above; objects in the dense Galactic bulge with more than 10,000 candidates were not evaluated, four objects could not be searched (a degenerate line of variations, and a damaged catalogue tile), and for most objects only one archival detection per season has been checked on images. The error rate of the archival identifications rests on the assumptions stated in Section~\ref{sec:archival}, and the tiers on nights of the catalogue hit, not all of which were checked on images. The magnitudes of the archival detections are on the Gaia $G$ scale with no colour term.

%==========================================================================================
\section{Availability}\label{sec:avail}
\os\ is written in Python with optional Rust kernels. The repository is private at the time of writing; the code, including the Rubin adapter used here (\cmd{orbitsift.rubin}), is available from \repourl. The following files accompany this paper as machine-readable supplements: \texttt{table2\_alerts.csv} (Table~\ref{tab:alerts} in full); \texttt{rubin\_alert\_associations.csv} (the \nRowsAll\ unassociated alert detections of designated objects: diaSourceId, designation, time, offset from the prediction and flags); \texttt{table1\_archival.csv} (Table~\ref{tab:archival} in full, both tiers); \texttt{table1\_not\_confirmed.csv} (objects whose archival detections passed the image check but which are not counted, with the reason); \texttt{earlier\_designations.csv} (the joint fits with earlier designations); \texttt{archival\_positions.csv} (the \posRows\ archival detections used by the fits: time, position, uncertainty, $G$-scale magnitude, filter, $S/N$ and exposure; obsTime is DATE-OBS plus half of EXPTIME, checked against the image header); \texttt{archival\_positions\_ades.psv} (the same detections in ADES format); and \texttt{decam\_proposals.txt} (the NOIRLab proposals of the DECam exposures used). The alert detections are Rubin's observations; they are listed for reference and are not submitted to the MPC by me. The archival detections are my measurements of public images and have not been submitted to the MPC; the same \posRows\ detections are given in ADES format (\texttt{archival\_positions\_ades.psv}, measurer G.~Cerati, validated with the IAU ADES tools) and will be submitted through the MPC's route for archival data after its review. No position of an undesignated object is given.

\begin{acknowledgments}
This paper uses public alerts of the NSF--DOE Vera C. Rubin Observatory (Alert Production, \citealt{rubinap}; Prompt Products, \citealt{rubinppdoi}; LSSTCam, \citealt{lsstcam}), produced by the LSST Science Pipelines \citep{rubinpipelines}. This material is based upon work supported in part by the National Science Foundation through Cooperative Agreements AST-1258333 and AST-2241526 and Cooperative Support Agreements AST-1202910 and 2211468 managed by the Association of Universities for Research in Astronomy (AURA), and the Department of Energy under Contract No.\ DE-AC02-76SF00515 with the SLAC National Accelerator Laboratory managed by Stanford University. Additional Rubin Observatory funding comes from private donations, grants to universities, and in-kind support from LSST-DA Institutional Members.

This work made use of the Fink resources. Fink is supported by LSST-France and CNRS/IN2P3. It uses the orbit catalogue and astrometry of the Minor Planet Center.

This research uses services or data provided by the Astro Data Lab \citep{fitzpatrick2014,nikutta2020}, which is part of the Community Science and Data Center (CSDC) Program of NSF NOIRLab, including the NOIRLab Source Catalog DR2 \citep{nidever2021}. This research draws upon DECam data as distributed by the Astro Data Archive at NSF NOIRLab. NOIRLab is managed by the Association of Universities for Research in Astronomy (AURA) under a cooperative agreement with the U.S.\ National Science Foundation. This project used data obtained with the Dark Energy Camera (DECam), which was constructed by the Dark Energy Survey (DES) collaboration. Funding for the DES Projects has been provided by the US Department of Energy, the US National Science Foundation, the Ministry of Science and Education of Spain, the Science and Technology Facilities Council of the United Kingdom, the Higher Education Funding Council for England, the National Center for Supercomputing Applications at the University of Illinois at Urbana-Champaign, the Kavli Institute for Cosmological Physics at the University of Chicago, Center for Cosmology and Astro-Particle Physics at the Ohio State University, the Mitchell Institute for Fundamental Physics and Astronomy at Texas A\&M University, Financiadora de Estudos e Projetos, Funda\c{c}\~ao Carlos Chagas Filho de Amparo \`a Pesquisa do Estado do Rio de Janeiro, Conselho Nacional de Desenvolvimento Cient\'ifico e Tecnol\'ogico and the Minist\'erio da Ci\^encia, Tecnologia e Inova\c{c}\~ao, the Deutsche Forschungsgemeinschaft and the Collaborating Institutions in the Dark Energy Survey. The Collaborating Institutions are Argonne National Laboratory, the University of California at Santa Cruz, the University of Cambridge, Centro de Investigaciones En\'ergeticas, Medioambientales y Tecnol\'ogicas--Madrid, the University of Chicago, University College London, the DES-Brazil Consortium, the University of Edinburgh, the Eidgen\"ossische Technische Hochschule (ETH) Z\"urich, Fermi National Accelerator Laboratory, the University of Illinois at Urbana-Champaign, the Institut de Ci\`encies de l'Espai (IEEC/CSIC), the Institut de F\'isica d'Altes Energies, Lawrence Berkeley National Laboratory, the Ludwig-Maximilians Universit\"at M\"unchen and the associated Excellence Cluster Universe, the University of Michigan, NSF NOIRLab, the University of Nottingham, the Ohio State University, the OzDES Membership Consortium, the University of Pennsylvania, the University of Portsmouth, SLAC National Accelerator Laboratory, Stanford University, the University of Sussex, and Texas A\&M University. Based on observations at NSF Cerro Tololo Inter-American Observatory, a program of NOIRLab (the \nProps\ proposal identifiers are listed in \texttt{decam\_proposals.txt}), which is managed by the Association of Universities for Research in Astronomy (AURA) under a cooperative agreement with the U.S.\ National Science Foundation. I thank Bill Gray for Find\_Orb.

\emph{Use of AI.} I conceived and designed the method, the search and the tests; the code, the analyses and the text were written with the assistance of an AI coding agent, Claude Code (Anthropic), under my direction. I read all of the code and the text and take full responsibility for the content of this paper.
\end{acknowledgments}

\facilities{Rubin:Simonyi, Blanco (DECam), Astro Data Lab}

\software{\os\ (Section~\ref{sec:avail}; \citealt{cerati2026survey}), Astropy \citep{astropy2022}, NumPy \citep{harris2020}, SciPy \citep{virtanen2020}, Matplotlib \citep{hunter2007}, jplephem, Find\_Orb \citep{findorb}}

\bibliographystyle{aasjournalv7}
\bibliography{refs}
\end{document}